\documentclass[5p,twocolumn,times,number]{elsarticle}

\usepackage{amssymb}

\begin{document}

\begin{frontmatter}

\title{Scintillating bolometers for Double Beta Decay search}

\author[UniMib,INFNmib]{L.Gironi\corref{cor1}}
\ead{luca.gironi@lngs.infn.it}

\cortext[cor1]{Corresponding author: tel: +39 02 64 48 21 07; fax: +39 02 64 48 24 63}

\address[UniMib]{Dipartimento di Fisica - Universit\`{a} di Milano Bicocca, Italy}
\address[INFNmib]{INFN - Milano Bicocca, Italy}


\begin{abstract}

In the field of Double Beta Decay (DBD) searches, the use of high resolution detectors in which background can be actively discriminated is very appealing. Scintillating bolometers containing a Double Beta Decay emitter can largely fulfill this very interesting possibility.
In this paper we present the latest results obtained with CdWO$_{4}$ and CaMoO$_{4}$
crystals. Moreover we report, for the first time, a very interesting feature of CaMoO$_{4}$ bolometers: the possibility to discriminate $\beta$ - $\gamma$ events from those induced by $\alpha$ particles thanks to different thermal pulse shape.

\end{abstract}


\end{frontmatter}

\section{Introduction}
\label{Intro}

The experimental evidence for neutrino oscillations clearly showed that the neutrino is a finite-mass particle. Anyway, two big questions concerning the neutrino are still unsolved: its nature (Dirac or Majorana) and the absolute value of its mass. Neutrinoless Double Beta Decay  (0$\nu$DBD) is, at present, one of the most sensitive method to study neutrino properties \cite{DBD} and bolometers are - together with germanium diodes - the detectors 
which have provided the best results so far. Bolometers allows the application of 
the so-called calorimetric approach, where the detector is composed of the same material candidate to the decay, and in the mean time allow the study of many isotopes by means of high resolution devices (FWHM around 0.2-0.5 \% above 2500 keV), this latter feature being necessary to resolve the searched peak from background.

The purpose of the so-called 0$\nu$DBD new generation experiments is to reach a sensitivity on the mass of the neutrino of the order of 50 meV, approaching - but probably not completely covering - the region of the inverse hierarchy \cite{CUORE,GERDA,MAJORANA,EXO}.
The high sensitivity required imply excellent energy resolution, a low number of spurious counts within the region of interest (low background) and a high quantity of the isotope on which the study focuses (large mass).

Further improvements cannot rely simply on the mass increase or on a better energy resolution but will require the implementation of innovative techniques for background discrimination.

In the case of a scintillating bolometer the double independent read-out (heat and scintillation) will allow, thanks to the different scintillation Quenching Factor between $\alpha$ and $\beta$ - $\gamma$, the suppression of $\alpha$ background events. These have been identified as the most important background source in bolometers dedicated to Double Beta Decay searches \cite{Arna, CUORE}.  Furthermore, the contribution of environmental gammas can be strongly reduced using a scintillating bolometer containing a DBD emitter whose transition energy exceeds the highest gamma line due to natural radioactivity (the 2615 keV line of $^{208}$Tl). Examples are $^{116}$Cd (Q$_{\beta\beta}$ $=$ 2805 keV), $^{82}$Se (Q$_{\beta\beta}$ $=$ 3000 keV) and $^{100}$Mo (Q$_{\beta\beta}$ $=$ 3030 keV). 
Finally, this technique is also extremely helpful for rejecting other unavoidable sources of background such as direct interactions of neutrons. 

\section{Environmental Background}
\label{EnvBack}

In a bolometric experiment, the signature of the neutrinoless Double Beta Decay (0$\nu$DBD) is a peak at the Q$_{\beta\beta}$ value of the transition, while in ordinary double beta decay (2$\nu$DBD) is a continuous beta-like spectrum. However, thanks to the excellent energy resolutions of bolometers, 2$\nu$DBD is seldom a significant source of background.

However, there are various sources that give rise to spurious counts in the region of interest such as external $\gamma$ background, neutrons, surface contaminations, $^{238}$U and $^{232}$Th internal contaminations and cosmogenic activity.
As already discussed, a way to eliminate the problem of $\gamma$ lines is to study isotopes with transition energy above the 2615 keV line. This is the $^{208}$Tl line that is the highest energy $\gamma$-ray line from natural radioactivity. Above this energy there are only extremely rare high energy $\gamma$ from $^{214}$Bi: the total Branching Ratio in the energy window from 2615 up to 3270 keV is 0.15 \% in the $^{238}$U decay chain. All the Double Beta Decay isotopes with Q$_{\beta\beta}>$2615 keV are listed in Table \ref{tab:isotopes}.

\begin{table}[]
\begin{center}
\caption{Double beta decay isotopes with endpoint energies above the $^{208}$Tl line.}
\begin{tabular}{ccc}
\hline\noalign{\smallskip}
\space Isotope   \space   & \space Q$_{\beta\beta}$ [MeV] \space & \space natural abundance \space \\
\noalign{\smallskip}\hline\noalign{\smallskip}
$^{116}$Cd & 2.80  & 7.5 \% \\
$^{82}$Se  & 3.00  & 9.2 \%\\
$^{100}$Mo & 3.03  & 9.6 \%\\
$^{96}$Zr  & 3.35  & 2.8 \%\\
$^{150}$Nd & 3.37  & 5.6 \%\\
$^{48}$Ca  & 4.27  & 0.19 \%\\
\noalign{\smallskip}\hline
\end{tabular}
\label{tab:isotopes}
\end{center}
\end{table}

Surface contaminations represents the main source of background for bolometric detectors such as CUORICINO \cite{Arna}. This source of background plays a role in almost all detectors but turns out to be crucial for fully active detectors, as in the case of bolometers: a radioactive nucleus located within a few $\mu$m of a surface facing the detector can emit an $\alpha$ particle (the energy of which is, in most of the cases, between 4 and 8 MeV). Those particles can lose part, or even all, of their energy in the few microns of this dead layer before reaching the bolometer. The energy spectra read by the bolometer, therefore, will result in a continuum between 0 and 4-8 MeV, covering, unfortunately all the possible Q$_{\beta\beta}$ values. Furthermore, the same mechanism holds in the case of surface contaminations on the bolometer itself.

\section{Scintillating Bolometer and Light Detectors}
\label{ScinBol}

\begin{figure}
\begin{center}
\includegraphics[ width=.7\linewidth]{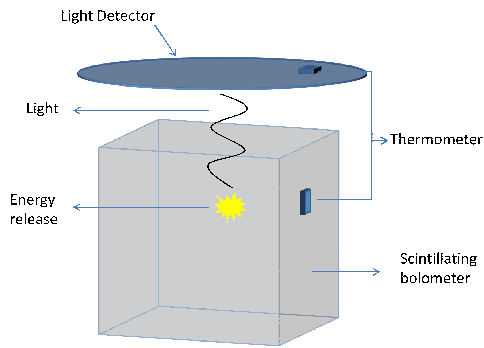}
\end{center}
\caption{Operating principle of scintillating bolometers. The release of energy inside a scintillating crystal follows two channels: light production and thermal excitation. The heat is read out by a temperature sensor (NTD) glued on the primary crystal while the light is read by a second crystal (light detector) where it is completely converted into heat.}
\label{bol_scint}
\end{figure}

A scintillating bolometer \cite{CaF2} is, in principle, a very simple device. It is composed of a calorimetric mass or absorber (a scintillating crystal coupled with a thermometer) and a Light Detector, facing it, and able to measure the emitted photons (see Fig. \ref{bol_scint}). The driving idea of this hybrid detector is to combine the two information available: the heat (i.e. that fraction of the energy released in the crystal absorber which is converted into phonons) and the emitted scintillation light (i.e. that small fraction of the energy which is converted into photons). Thanks to the different scintillation yield (or scintillation Quenching Factor) of different particles (namely $\beta$ - $\gamma$, $\alpha$ and neutrons) they can be very efficiently discriminated.

The idea to use  a bolometer as light detector was first developed by Bobin et al. \cite{Coron97} and then optimized \cite{CRESST,ROSEBUD} for Dark Matter (DM) searches. Starting from that work we developed a thermal light detector to be used for DBD search.

Our light detector is constituted of a ``dark'' thin bolometer, normally a crystal of Ge or Si, that can absorb scintillation photons and give a measurable thermal signal. This means that it has the characteristic time constant of bolometers (20 - 500 ms) and have an important advantages respect PMTs: they are sensitive over an extremely large band of photon wavelength.

\section{Experimental setup and Results}
\label{ExperSet}

In these years different crystals were tested in dedicated runs to study their characteristics such as their thermal response, light yield and radio purity. The last test was performed at the beginning of 2009 in order to evaluate the efficiency of discrimination of this technique. In this test were measured some new CdWO$_4$ and CaMoO$_4$ crystals.

The detectors are mounted in  an Oxford 200 $^{3}$He/$^{4}$He dilution refrigerator located deep underground in the National Laboratory of Gran Sasso. The cryostat is surrounded by about 20 cm of lead in order to reduce environmental $\gamma$ radioactivity. Furthermore, the crystal set-up was mounted below about 5.5 cm of Roman lead in order to further decrease the environmental background. The cryostat is also surrounded by about 7 cm of polyethylene (CH$_2$) to thermalize fast neutrons and about 1 cm of CB$_4$ that, thanks to the high neutron capture cross section for thermal neutrons of $^{10}$B, allows to reduce the neutron flux on the detectors.

The temperature sensors are Neutron Transmutation Doped (NTD) Ge thermistors glued to each crystal. A Si resistor of  $\sim$300 k$\Omega$ is attached to each crystal and acts as a heater to stabilize the gain of the bolometer \cite{ALES98}. The read-out of the thermistors is performed via a cold ($\sim$110 K) preamplifier stage located inside the cryostat \cite{EF}. The room temperature front-end \cite{programmable} and the second stage of amplification are located on the top of the cryostat. After the second stage there is an antialiasing filter (a 6 pole roll-off active Bessel filter). The ADC is a NI USB device (16 bit 40 differential input channels) located in a small Faraday cage. The signals (software triggered) are sampled in a 512 ms window with a sampling rate of 2 kHz.

The data analysis is completely performed off-line. It uses an Optimal Filter technique \cite{CUORE} to evaluate the pulse amplitudes and to compare pulse-shapes with detector response function. Events not caused by interactions in the crystals are recognized and rejected on the basis of this comparison. Pile-up pulses are identified and dealt with. Accepted-pulse amplitudes are then corrected using the variation in the gain measured with the heat pulses from the Si resistors.

Since, for bolometric detectors, surface alpha contaminations represents the main source of background in DBD region \cite{Arna}, in this run $^{238}$U sources was faced to the crystals. The purpose of these sources is to have a high number of alpha counts in the 2-3 MeV region in order to evaluate the efficiency of discrimination of this technique. For this reason, after simulations with the GEANT4 \cite{Geant4} package and measurements with a Silicon Barrier Detectors, alpha sources were covered with an aluminated Mylar foil (thickness $\sim$6 $\mu$m). In such a way alpha particles lose part of their energy in this dead layer before reaching the bolometer. The energy spectra read by the bolometer, therefore, will result in a continuum between 0 and $\sim$4 MeV.

\subsection{CdWO$_{4}$}
\label{CdWO4}

In 2008 we tested an array of four CdWO$_{4}$ Double Beta Decay scintillating bolometers, read by only one light detector \cite{Gironi08}. The array consists of a plane of four 3x3x3 cm$^3$ crystals and a second plane consisting of a single 3x3x6 cm$^3$ crystal. This test demonstrate the technical feasibility of this technique through an array of detectors, and perform a long background measurement in the best conditions in order to prove the achievable background in the 0$\nu$DBD region.

\begin{figure}
\begin{center}
\includegraphics[ width=1.\linewidth]{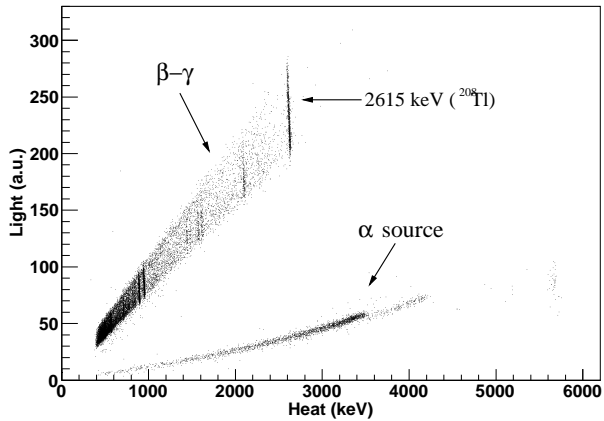}
\end{center}
\caption{Scatter plot of Heat (new CdWO$_4$ crystal, h = 50mm, $\varnothing$ = 40mm, 508g) vs Light. Calibration performed on $\gamma$ peaks.}
\label{cdwo4}
\end{figure}

In 2009 a measurement was done with a new, bigger, cylindrical CdWO$_{4}$ crystal (h = 50mm, $\varnothing$ = 40mm, 508g). The main difference between this crystal and the crystals measured previously is that this crystal has perfect optical surfaces. This feature has led to a significant reduction of the width of the alpha band compared to measurements made previously. In figure \ref{cdwo4} it is possible to observe the high number of counts in the region 2-4 MeV due to alpha source facing the detector. Data relates to a measurements made with the new crystal with a live time of $\sim$ 93 h. This plot shows how it is possible to separate very well the background due to $\alpha$ particles from the $\beta$ - $\gamma$ region. In particular it should be noted that $\alpha$ continuum is completely ruled out thanks to the combined measurement of heat and scintillation.

The resolution of the main bolometer is affected by the fluctuation of the
fraction of energy going in the heat and light channels. Since the energies of the
two channel are anti-correlated, it is possible reconstruct the intrinsic energy resolution
of the bolometer. Rotation the axis by $\theta$ we obtain a new plot in which
E$_{rot}$ = E$_{heat}$ $ cos \theta$ + E$_{light}$ $ sin \theta$ is the new energy. After rotation, the energy resolution at the 2615 keV line is 12 keV FWHM.

\subsection{CaMoO$_{4}$}
\label{CaMoO4}

Recently CaMoO$_{4}$ was intensively studied as possible cryogenic scintillation bolometric
detector for experiments to search for DBD and Dark Matter \cite{Seny06,Mikh06-JPDAP,Pirr06}. Below we discuss possible use of CaMoO$_4$ scintillators as potential detectors in search for 0$\nu$DBD decay of $^{100}$Mo.

In the case of scintillating bolometer, thanks to the excellent energy resolution achievable respect to scintillators, the 2$\nu$DBD of $^{100}$Mo will not contribute to the 0$\nu$DBD peak. Nevertheless the background due to 2$\nu$DBD of $^{48}$Ca will limit the reachable sensitivity. In fact contribution from 2$\nu$DBD of $^{48}$Ca could be very dangerous (measured half-life of this process is $T_{1/2}(2\nu)=4\times10^{19}$ yr \cite{Tret}). $^{48}$Ca present in natural Ca (a.i.=0.187\%) produces a background of about 10$^{-3}$ counts/keV/kg/y in the 0$\nu$DBD region of $^{100}$Mo.

Even if sensitivity of CaMoO$_{4}$ crystal is limited by 2$\nu$DBD decay of $^{48}$Ca, this crystals shows, for the first time, the possibility to discriminate $\beta$ - $\gamma$ from $\alpha$ thanks to different shape of the thermal pulses.

\begin{figure}
\begin{center}
\includegraphics[ width=1.\linewidth]{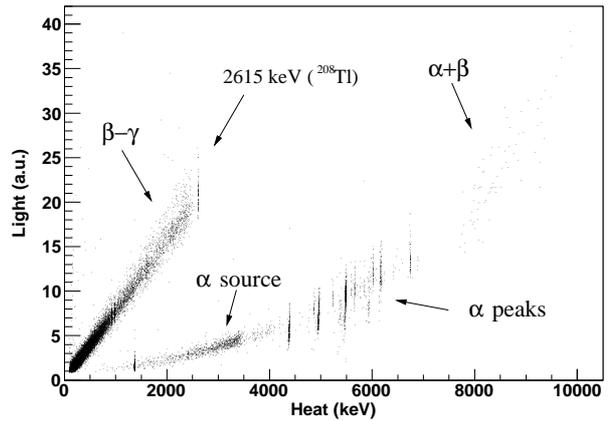}
\end{center}
\caption{Scatter plot of Heat (CaMoO$_4$ crystal, h = 40mm, $\varnothing$ = 35mm, 158g) vs Light. Calibration performed on $\gamma$ peaks.}
\label{fig:camoo4}
\end{figure}

\subsubsection{Bolometric Pulse Shape Analysis}
\label{PSA}

The release of energy inside a scintillating crystal follows two channels: light production and thermal excitation. The heat is read out by a temperature sensor glued on the primary crystal while the light is read by a second crystal (light detector) where it is completely converted into heat. In this run the CaMoO$_{4}$ crystal was measured with an antialiasing filter with a high cutoff frequency of 120 Hz. This feature allow us to study in detail the rise time of the thermal signal in the CaMoO$_{4}$ crystal. 

The light pulses produced by $\alpha$ particles and $\beta$ - $\gamma$ in the
CaMoO$_{4}$ scintillator present different decay time constants and relative intensities. Moreover it was observed a temperature dependence of averaged decay time of the light pulses \cite{Anne08}. This difference between the light pulses and the increase of the decay time costant of the light pulses at low temperature modify the shape of the thermal signal measured on the same scintillating crystal.

\begin{figure}
\begin{center}
\includegraphics[ width=1.\linewidth]{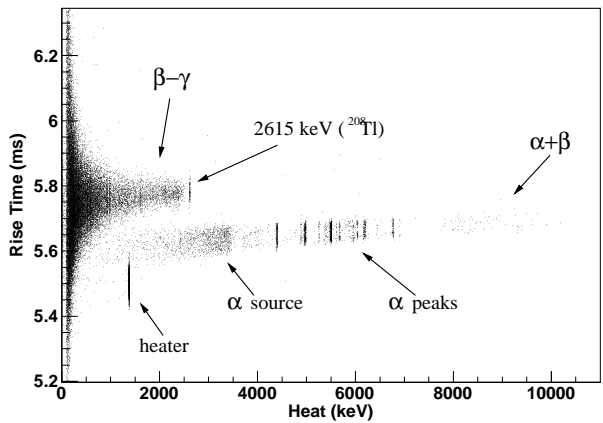}
\end{center}
\caption{Scatter plot of Heat vs Rise Time in CaMoO$_{4}$ crystal.}
\label{fig:camoo4_RT}
\end{figure}

As can be seen in figure \ref{fig:camoo4_RT} the difference in the rise time of the thermal pulses measured in the scintillating crystal allows to discriminate $\beta$ - $\gamma$ events from those induced by $\alpha$ particles. In this run the efficiency of discriminations of $\alpha$ events in the 2.7 - 8 MeV region is 99.7\% and can be improved by measurements devoted to study this feature.

The possibility to discriminate $\alpha$ events thanks to the different pulse shape allow to consider an experiment without light detector and reflecting sheets. This will allow not only to construct a 0$\nu$DBD experiment in a much easier way but also to develop a new technique in order to investigate with a very high sensitivity surface contaminations. Indeed the surface of a scintillating bolometer that can be faced to a sample is much bigger than the surface of a standard Silicon Barrier Detector. Moreover scintillating bolometers have not dead layer and can reach energy resolution orders of magnitude better than standard Silicon Barrier Detector.

\section{Conclusion}

We have performed  simultaneous detection of heat and light  on different Double Beta Decay scintillating crystals. Very good results was obtained with CdWO$_4$ and CaMoO$_{4}$ crystals. CdWO$_4$ has excellent characteristics that make it an excellent candidate as a crystal for the study of 0$\nu$DBD. 

The sensitivity of CaMoO$_{4}$ is limited by the background due to 2$\nu$DBD of $^{48}$Ca. However, this crystal shows, for the first time, the possibility to discriminate $\alpha$ events through the study of the shape of the thermal signal. This feature can be used to develop detectors for measurements of surface contamination with high sensitivity.

\section{Acknowledgments}

This work was funded and implemented under the Bolux experiment of INFN.

\end{document}